\documentclass[preprint]{aastex}

\newcommand \degpoint {${.}\!\!^{\circ}\!$}
\newcommand \secpoint {${.}\!\!^{\prime \prime}\!$}
\newcommand \minpoint {${.}\!\!^{\prime}$}

\shorttitle{Outer Halo of M31}
\shortauthors{Durrell et al.}

\begin{document}

\title{Photometry and the Metallicity Distribution of the Outer Halo of M31. \\
II. The 30 Kpc Field}

\author{Patrick R. Durrell\altaffilmark{1}}

\affil{Department of Astronomy \& Astrophysics, The Pennsylvania State
University, 525 Davey Lab, University Park, PA  16802  USA}
\email{pdurrell@astro.psu.edu}

\author{William E. Harris\altaffilmark{1}}
\affil{Department of Physics \& Astronomy, McMaster University, Hamilton, ON  L8S 4M1  Canada}
\email{harris@physics.mcmaster.ca}

\and

\author{Christopher J. Pritchet}
\affil{Department of Physics \& Astronomy, University of Victoria, Victoria, BC
V8W 3P6  Canada} \email{pritchet@uvic.ca}

\altaffiltext{1}{Visiting Astronomer, Canada-France-Hawaii Telescope, operated by the
National Research Council of Canada, le Centre National de la Recherche Scientifique
de France, and the University of Hawaii}

\begin{abstract}

We present the results of a wide-field $(V,I)$ photometric study of the
red-giant branch (RGB) stars in the outer halo of M31, in a field located 30 to
35 kpc from the center of the galaxy along the southeast minor axis.  At this
remote location, we find that RGB stars belonging to M31 are sparsely but
definitely present, after statistical subtraction of field contamination. We
derive the metallicity distribution (MDF) for the halo stars using
interpolation within a standard $(I,V-I)$ grid of RGB evolutionary tracks. The
halo MDF is quite broad but dominated by a moderately high-metallicity
population peaking at [m/H]$\sim -0.5$, strikingly different from the [m/H]
$\sim -1.3$ level which characterizes the outer halo of the Milky Way. However,
the shape and peak metallicity for this region are entirely similar to those
found in other studies for the inner regions of the M31 halo, particularly our
previous study of a 20-kpc region (Durrell, Harris, \& Pritchet 2001) employing
similar data.  In summary, we find no evidence for a metallicity gradient or
systematic change in the MDF out to quite large distances in the M31 halo: it
appears to be a homogeneous and moderately metal-rich subsystem of the galaxy
at all locations.   The star counts in the 30-kpc field are also consistent
with the $r^{1/4}$ law that fits the interior regions of the M31 spheroid
surface brightness profile.    The metal-rich MDF and the $r^{1/4}$ spheroid
suggests M31 more strongly resembles a giant elliptical galaxy than other,
Milky-Way-like, spirals.

\end{abstract}


\keywords{galaxies: halos --- galaxies: individual (M31) --- galaxies:
photometry--- galaxies: stellar content --- Local Group}

\section{Introduction}

The halos of galaxies provide
us with a fossil record of the earliest processes of galaxy formation.
However, studying the halo content is particularly difficult in
disk and spiral galaxies, where the halos are sparsely
populated when compared to their disk
subsystems.  Therefore observations of halos in
disk galaxies require either extremely deep surface photometry studies
\citep[eg.][]{wu02,zib03}, or photometric and kinematic studies of the individual
stars, star clusters, or planetary nebulae that populate these halos.
Only in the nearest galaxies can the latter method(s) be employed.

The halo of the spiral galaxy M31 provides us with a uniquely large, nearby,
and readily accessible population of halo stars in a giant disk galaxy.  It is
near enough that photometry of its brightest red-giant branch (RGB) stars is
easily carried out with 4-meter-class telescopes.   Since the pioneering
color-magnitude diagram (CMD) of M31's halo stars by \citet{mk86}, a large
number of CMD-based studies have been carried out from the ground \citep{mou86,
pv88, ch91, dav93, dur94, cou95,reit98,dur01,ferg02} and using the Hubble Space
Telescope \citep{hol96, rich96a, rich96b, svd01, bell03}. These studies
consistently show that the majority of M31 halo stars are redder than their MW
counterparts, a difference usually attributed to higher metallicity. These
studies (most of which are based on fields located at projected
distances from $\sim 7$ to over 20 kpc
from M31, and most lying near the SE minor axis of M31) find the mean
metallicity for the M31 halo stars to be in the range $-1 <$ [Fe/H] $< -0.6$,
showing remarkable consistency for datasets differing greatly in photometric
depth, field size, and observed colors.  Whether or not the mean halo [Fe/H]
varies with radius is unclear, although \citet{vp92} suggested that a strong
metallicity gradient does {\it not} exist in the inner halo, and recent studies
in more remote fields suggest that this is indeed the case
\citep{reit98,dur01,ferg02}.  Furthermore, these studies show that the M31 halo
stars cover a wide range of metallicity ($-2 \lesssim $[Fe/H]$ \lesssim 0$), a
conclusion further confirmed by spectroscopic studies \citep{reit02}.   For
comparison, the MW halo is more metal-poor by nearly an order of magnitude in
the mean ([Fe/H]$\sim -1.6$; Ryan \& Norris 1991).

In addition to the differences in the metallicity distribution functions (MDFs), 
the spatial density of halo stars
in the MW follows a power law \citep{wm96,iv03}, which differs from the
$r^{1/4}$ profile of the M31 bulge/halo found by \citet{pv94}, more like
that of an elliptical galaxy.
Interestingly, the MDF of the M31 halo shows
remarkable similarities with that of the giant elliptical galaxy NGC 5128 and
the compact elliptical M32 \citep{dur01,hh01,hh02}; the high mean metallicity
of the latter \citep{gri96} has led to the suggestion that perhaps tidal
disruption of M32 \citep[eg.][]{choi02} has helped create the largely metal-rich M31
halo \citep{ferg02}.   There are a number of other scenarios envisioned for
matching current observational results of the M31 halo, including a massive
merger \citep{brown03}, formation via accretion of smaller, stellar fragments
\citep{cote00}, and in situ star formation from the agglomeration of smaller,
primarily gas-rich objects \citep{dur01}.

Recently, there have been exciting new revelations on the structure of
the M31 halo.  \citet{ib01} have reported the detection of a large
stream of metal-rich stars SE of M31 in their wide-field survey, and
\citet{ferg02} showed that there is more complex substructure in the
halo and outer disk regions of M31.  \citet{mor03} and
\citet{zucker04} have also found evidence for additional tidally disrupted
satellite galaxies in the M31 halo.  In an observational {\it tour de force},
\citet{brown03} have created an extremely deep HST-based CMD of the
M31 halo about 11 kpc from the nucleus, and have confirmed the basic
results of previous work (largely metal-rich halo, fewer metal-poor
stars).  However, they also deduce that the metal-rich stars may be a
few Gyr younger than that of the (old) metal-poor stars and suggest
that this age difference originates from a major merger 6-10 Gyr ago
(the metal-rich population requiring incorporation of a rather massive
galaxy).  These are all interesting possibilities that require further
study, and indicate that the M31 halo is not simply a homogeneous,
coeval population of metal-rich stars.

\section{Observations + Data Reduction}

In our first paper (Durrell et al 2001; hereafter Paper I),
we described $VI$ photometry for over 2000 RGB halo stars in a
halo field located about 20 kpc from M31 along the southeast minor axis.
We found that the MDF at this projected location is, like the inner halo,
still strongly weighted towards more metal-rich stars with a major peak
in the MDF at [Fe/H]$\sim -0.8$.  The MDF also has a well populated tail
extending to lower metallicities, making the entire distribution extremely
broad.\footnote{We note that the metallicities in Paper I and this paper
are typically quoted as [m/H], where [Fe/H] $\sim$ [m/H] $-$ 0.3 [assuming an
enhanced [O/Fe]=$+0.3$].}   In the discussion of Paper I
we suggested that the entire halo
MDF could be matched by a classic ``simple model'' of chemical evolution
with wind-driven mass loss, and
could be explained by a formation scenario where much of the M31 halo formed
though the accretion of primarily gas-rich fragments.

In the present paper, we describe new results for a field located $\sim 30$ kpc
from M31 also along the SE minor axis, and create the first clearly defined
halo\footnote{The G1 field studied by \citet{rich96a} and \citet{bell03} is
located $\sim 35$ kpc from M31's nucleus, but it is clear this area is
contaminated by disk stars, and may also have tidal streams/features
\citep[eg.][]{reit03,rich03}.   This field does not represent a `clean' halo
field at $r=35$ kpc.} MDF at such a large distance from M31.

The observations we present here are of a single field centered at $\alpha_{2000} = 0^h
51^m 09^s$, $\delta_{2000} = +39^{\circ} 46^{\prime} 40^{\prime \prime}$,
the field referred to as $\mathcal{M}3$ by \citet{pv94}.   We
used the UH8K camera at the Canada-France-Hawaii Telescope (CFHT) on Sept. 19,
1996 to acquire the data, during the same observing run from which our Paper I
observations were obtained.  The UH8K camera (now decommissioned) was an 8-CCD
mosaic camera with a total imaging area of 8192$^2$ pixels or 28\minpoint 1
$\times$ 28\minpoint1 (scale = 0\secpoint 206 per pixel).  One of the chips
(chip 4) showed considerable bleeding and other inherent structure, and the low
QE of chip 6 also proved to make it of little use for our purposes. We have
thus used the remaining 6 chips (covering an area of 21$^{\prime}$ x
28$^{\prime}$ on the sky) for all subsequent analysis.  The center of our field
is located 2\degpoint 2 SE of the M31 nucleus along the minor axis, or at a
projected distance of $\sim 30$ kpc from M31, assuming $d=780$ kpc for M31
\citep{hol98, dur01}; the entire field samples stars located from 1\degpoint 90
to 2\degpoint 5 (27 to 35 kpc) from M31.  The location of the $\mathcal{M}3$
field is plotted in Figure 1, which also illustrates the fields from other
recent M31 halo studies (including our $\mathcal{M}2$ [20 kpc] field from Paper
I).  Our field $\mathcal{M}3$ is located just within the outer extent of the
M31 halo survey of \citet{ib01}.

\subsection{Calibration and Pre-processing}

Our science images of the $\mathcal{M}3$ field were obtained under
non-photometric conditions, although shorter exposure images
of the field were taken on the
(photometric) night of Sept. 21, 1996 during the same observing run which we
could use to calibrate our data.   The
photometry was calibrated through \citet{lan92} standards observed during the
entire observing run (Sept 19-23, 1996).   The calibration equations were
derived for stars on a single chip (chip 1) of the UH8K Mosaic, and zeropoint
shifts ($\Delta V_{1,n}$, $\Delta I_{1,n}$) were derived to make the
appropriate corrections to the $V$ and $I$ magnitudes measured on the other
chips (a detailed description can be found in paper I):

\begin{equation}
 V = v_n - 0.152 {\rm X} - 0.034{\rm (V - I)} + z({V_1}) + \Delta V_{1,n}
\end{equation}
\begin{equation}
  I = i_n - 0.061 {\rm X} + z({I_1}) + \Delta I_{1,n}
\end{equation}

\noindent Here $X$ is the mean airmass (the airmass terms are adopted values
from Landolt 1992), $z(V_1)$ and $z(I_1)$ are the zeropoint values derived for
chip 1, and $v_n$ and $i_n$ are the instrumental aperture magnitudes for
chip $n$, as measured through an aperture
radius of 3.3$^{\prime\prime}$ and normalized to an exposure time of 1 second.
The rms scatter of the standard stars
in each of the calibration equations was 0.018 -- 0.027 magnitudes.

As with the previous data presented in Paper I (the inner $\mathcal{M}2$ field
and a very remote background field $\mathcal{R}1$), the exposure times were 4 $\times$
900s in $V$ and 3 $\times$ 900s in $I$, with seeing on most individual images
of $\sim$0\secpoint 8 to 1\secpoint 0 (FWHM).

The program images were pre-processed with bias frames, dark frames and
flat-fields, and then combined in the normal manner through IRAF\footnote{IRAF
is distributed by the National Optical Astronomy Observatories, which are
operated by the Association of Universities for Research in Astronomy, Inc.,
under cooperative agreement with the National Science Foundation.} tools.  Flat
fielding was carried out on the individual chips, using either a combined
`superflat' or a twilight flat that had been further corrected using a heavily
smoothed superflat.   The resulting images had uniform background intensities
to less than $\simeq 1\%$.  The individual CCD images were re-registered,
scaled and averaged to construct a single combined image for each chip in each
filter.  The FWHM of the stellar images in all of the final, combined images
was $\sim$ 0\secpoint 90 for the $V$ images, and $\sim$0\secpoint 95 for the
$I$ images.

\subsection{Photometry and Image Classification}

   Photometry of the objects on each combined image was performed
with the stand-alone versions of the DAOPHOT II / ALLSTAR packages
\citep{stet87, stet90, stet92}.  A single pass of DAOPHOT II $+$ ALLSTAR with a
3.5$\sigma$ detection threshold was adopted, since tests showed that a second
pass of DAOPHOT II did not add noticeably to the number of objects detected on
these relatively uncrowded fields.   A stellar point-spread-function (PSF) was
derived from 17 to 32 bright, uncrowded stars per image.   A constant PSF was
found to adequately fit the data on all images.

We are only interested in objects on the images with stellar appearance, so any
resolved background galaxies that were not already ignored by DAOPHOT II were
subsequently removed with a combination of image parameters: the DAOPHOT $\chi$
parameter \citep{stet87}, and the $r_{-2}$ image moment from \citet{kron80} and
\citet{har91}\footnote{We add that we did not use one of the image classifiers
utilized in Paper I, namely $\Delta M$ (the difference between the PSF and
aperture magnitudes), because it eliminated few objects that were not already
rejected through either DAOPHOT $\chi$ or $r_{-2}$.}.  All objects exceeding
{\it either} $\chi_{max}$ or $r_{-2,max}$ were considered to be non-stellar and
rejected from further analysis. We used artificial-star experiments (see below)
to set the classification boundaries and to confirm that few genuinely stellar
objects were rejected.

\section{Color-Magnitude Diagram}

The separate $V$ and $I$ lists of stellar objects from each chip were
merged with a matching radius of 1 pixel, and then used to create $I,
(V-I)$ color-magnitude diagrams for each. The CMD of the entire
$\mathcal{M}3$ field (all 6 CCDs) is plotted in Figure 2, along with
the background field $\mathcal{R}1$ for comparison.  We note that
the CMDs from the individual mosaic chips have slightly different photometric
completeness limits (see next section) due to small QE differences,
and we show the extreme values for these limits in Figure 2.  Typical
errorbars (derived from the artificial star experiments; see below)
for stars with $(V-I) = 1.0$ are also plotted.

\subsection{Have We Detected the RGB Population?}

There are a total of 6292 starlike objects in the $\mathcal{M}3$ CMD,
roughly $\sim 35\%$ more than present in the background $\mathcal{R}1$
field (4648 stars; paper I).  Over the nearly 550 arcmin$^2$ in each field,
the density of stars is thus quite uncrowded in an absolute sense.
The usable field size of the $\mathcal{M}3$ field is 546 arcmin$^2$,
almost exactly the same size as the background field, so roughly
one-third of the stars in our outer-halo field should be M31 halo
stars.  Most of the brighter ($17 < I < 20.5$) objects are foreground
Milky Way halo stars, and most of the fainter, bluer objects are
unresolved background galaxies.

We have performed a series of tests to confirm that we are indeed
detecting the M31 halo this far from the galaxy center.   
First, we define a box within the CMD ($20.5 < I < 22.5, 1.0 < (V-I) < 2.5$)
within which we expect the upper part of the M31 RGB to be present, and
simply count the numbers of stars in each of our program fields that
lie within that box.  For $\mathcal{R}1$, there are 1034 such stars 
(assumed to be entirely background contamination);
for $\mathcal{M}3$, there are 1603 stars; and for $\mathcal{M}2$ -- our
field from Paper I -- there
are 3312 stars.  Since this part of the CMD is bright enough and blue enough
that all three fields should have similar photometric limits and 
high completeness, we can 
use these totals to make {\sl rough} estimates of the statistical
significance of the RGB detection.  From the count statistics (and for the
moment ignoring slight differences in the foreground reddening of each
field; see below), we then estimate that we detect $(569 \pm 51)$ upper-RGB
stars in field $\mathcal{M}3$, and $(2278 \pm 66)$ in the mid-halo field
$\mathcal{M}2$.  This test is suggestive that we are detecting the RGB
at high significance, even though the background contamination 
is $2 \times$ larger than the RGB ``signal'' in $\mathcal{M}3$.
A more rigorous version of the field subtraction is done below.

A slightly more advanced test is to ask whether this excess residual population
follows the luminosity function (LF) that would be expected for a true RGB.
We have constructed a smoothed LF of the residual population 
$\mathcal{M}3 - \mathcal{R}1$ in the same way as we did in Paper I.
The result is shown in Figure 3.  Here we use a
smoothing kernel of 0.03 mag, and to remove some of the most obvious
background contamination brighter than the RGB tip, we exclude stars
with $I < 20.5$ bluer than $(V-I)=1.4$ or redder than $(V-I)=3.0$
(again, the same procedure we used for field $\mathcal{M}2)$.
The residual LF is noisier than we found in Paper I, but there is a clear excess of
stellar objects fainter than $I=20.6$ with numbers rising towards
fainter magnitudes in just the same way as we found in Paper I (see for
comparison Fig.6 in that paper).
{\sl Brighter} than this level, the background population subtracts
cleanly away with no significant residuals.  The $\mathcal{M}3$ field
passes the LF consistency test.

As a third test, we have also performed a one-by-one statistical
subtraction of background stars from the $\mathcal{M}3$ CMD.
To do this, we take each star in the $\mathcal{M}3$ CMD list
and calculate the ``separation radius'' $r$ to every star in 
the $\mathcal{R}1$ list as 
$r=\sqrt{(0.2[I_{R1} - I_{M3}])^2 + ((V-I)_{R1} - (V-I)_{M3})^2}$.
If any star in $\mathcal{R}1$ has $r < 0.05$ mag, then both it and
its matching object in $\mathcal{M}3$ 
are flagged and removed from the lists.  In this way we match up stars
within a fairly narrow range in color but a more generous range in magnitude
(the distance, i.e.~apparent magnitude, of the background star is
less important for this purpose than its color, i.e.~population type).
This process is continued until every star in $\mathcal{M}3$ has been tried.

In Figure 4 we show, first, the CMD for the remaining 
{\sl unmatched} $\mathcal{M}3$ stars, and second, the CMD of all the
{\sl matched} $\mathcal{M}3$ stars.  By hypothesis, the first of these two
CMDs should be a nearly pure RGB population from the M31 halo, while
the second CMD should be a nearly pure background population (and should resemble
the original CMD of the $\mathcal{R}1$ field, from Fig.~2).
In Fig.~4a, an RGB stands out clearly in the expected region of the CMD,
while other areas have almost entirely subtracted away.
While this residual ``signal'' is much smaller than the population we found in
the $\mathcal{M}2$ (20 kpc) field, features of the M31 halo RGB are
still clear, such as the blue `edge' of stars at $(V - I) \sim 1.3$
and $I\sim 20.5-22.0$, which are the most metal-poor ([Fe/H]$<-1.5$) M31
halo stars.   

This statistically cleaned CMD does not take into account slight
place-to-place differences in limiting magnitude or photometric
completeness, and so we do not use it to do any direct analyses
of the metallicity distribution or number counts.  However, we
regard it as a very strong test of the genuine presence of the M31
halo in our outer field.

\subsection{Artificial Star Experiments}

The photometric
completeness $f$ of a star in the CMD is a function of $I$, $(V-I)$,
and also location within the CCD mosaic because of chip-to-chip differences
in QE within the array.
Here, $f$ is defined as $N_{rec} / N_{add}$, where
$N_{add}$ is the total number of stars added to the science images with
a particular value of $(I,V-I)$, and $N_{rec}$ is the total number of those
stars that are recovered, regardless of where they were found on the CMD.

To trace the completeness function behavior, we performed a series of
artificial-star experiments on each chip.
These tests are the same as those carried out in Paper I for other fields
in our survey.  A total of 40000 stars (10 runs of 4000 total stars added per
run) were added to the {\it star-subtracted} science images for each chip.  The
stars had discrete input magnitudes and colors sampled at 0.5 magnitude intervals
in the grid ($20.5 < I < 24.0$, $0.0 < (V-I) < 3.0$).
The simulated images were then reduced in precisely the same way as
described above -- a single pass of DAOPHOT II/ALLSTAR, merging of the
resulting $V$ and $I$ datasets, and removal of non-stellar images using the
image-classification algorithm.  The number of stars added was similar to those
found on the original science frames, indicating that we have adequately
re-created the rather low crowding conditions on the frames.

The limiting magnitudes ($m_{lim}$; defined as the 50\% completeness level) for
each individual chip were derived by fitting the interpolation
function \citep[see][]{fl95} :

\begin{equation}
 f(m) = {{1}\over{2}}\left( 1 - {{\alpha(m - m_{lim})}\over{\sqrt{1 + \alpha^2(m - m_{lim})^2}}} \right)
\end{equation}

\noindent where $\alpha$ is a parameter that
measures the rate of decline of $f(m)$ at $m_{lim}$.  The
values for $I_{lim}$ and $V_{lim}$
for each chip are listed in Table 1.   Our results can be compared directly to
those obtained in Paper I for the other fields in our survey; the limits of our
$\mathcal{M}3$ data are comparable to that of the background $\mathcal{R}1$
field, but different from the interior $\mathcal{M}2$ field (the $\mathcal{M}3$
limits are $\sim 0.5$ mag brighter in $I$, and $\sim 0.5$ deeper in $V$).  These
differences are due to the sky conditions; our $\mathcal{M}2$ data had better
overall seeing (0\secpoint7 vs. 0\secpoint9), but much brighter sky conditions
for the $V$ images.

Each star observed in our CMD (Figure 2) was then assigned a value of
$f(I,V-I)$ based on linear interpolation between the grid points from the
artificial star results from the chip the star was measured on; it is these
values that are used to derived corrected number counts for our MDF analyses
below.

\subsection{Reddening}

In Paper I, the appropriate total reddening $E(V-I)$ for the
$\mathcal{R}1$ and $\mathcal{M}2$ fields was directly determined from
the location of the blue edge of the foreground Milky Way halo
population in the CMDs.  In Figure 5 we have matched the color
histogram of all stars with $17 < I < 20$ (all, presumably, foreground
Milky Way halo stars) in the $\mathcal{M}3$ CMD with the derived blue
edge of the $\mathcal{M}2$ field at $(V-I)=0.64$.  To match the
histograms, we need to shift the $\mathcal{M}3$ data 0.03 mag to the
red, suggesting $E(V-I)=0.07\pm 0.03$ for this field (where we have
assumed an error of 0.02 mag on the color difference).  This observed
difference is due to the (slightly) higher galactic latitude of the
$\mathcal{M}3$ field, but is also affected by the uncertainties ($\sim
0.02$ mag) in the photometric color zeropoint; the latter is likely
the dominant factor.  This result is in excellent agreement with the
predicted value ($E(V-I)=0.07$) we find from the maps of
\citet{sch98}.  In addition, by adopting this empirical approach to
deriving $E(V-I)$ for a given field, we effectively remove any
systematic errors resulting from field-to-field color calibration
differences.

\section{Analysis}

\subsection{Metallicity Distribution Function}

In order to construct a metallicity distribution function (MDF) from
our CMD, we need to assume an age distribution for the halo stars.  To
do so, we have assumed that all the halo stars are comparably old, so
that by hypothesis the locations of the RGB loci are determined
primarily by metallicity.  While this may seem at odds with the
finding of a intermediate-age M31 halo population by \citet{brown03},
we note that this result (derived for a field located $\sim 11$ kpc
from M31) may not be applicable to our much more remote halo fields.
More importantly, the locus of the intrinsic RGB is insensitive to age
for stellar populations that are already old, and thus any large
spread in RGB location is most likely to be due to metallicity
differences.  In a more practical direction, since our data do not
reach the fainter but more age-sensitive regions of the CMD (the
horizontal branch and main-sequence turnoff), they {\it cannot} be
used to investigate the age distribution of the M31 halo without {\it
a priori} knowledge of the stellar metallicities that the RGB gives us
to first order.  Here, we simply assume the same age distribution
(effectively, the same set of fiducial RGB tracks) as used in Paper I
so that we can make the most direct possible comparison between the
two fields.  Clearly, in future it would be of great interest to find
out if the inner-halo age spread found by \citet{brown03} persists
into the outer halo, but even an age range of $\sim 5$ Gyr would not
strongly affect the MDF deduced from the RGB; for a star of a given
$(V-I)$ color, changing the age from 12 to 5 Gyr would increase the
derived metallicity by $\sim 0.2$ dex \citep[e.g.][]{hh99}.

In practice, we use stars in the brightest two magnitudes of the RGB
and a grid of fiducial RGB model tracks to estimate the abundance
[m/H] of each star (defined as [m/H] = log($Z/Z_{\odot}$), where we
adopt $Z_{\odot}=0.0172$ as in Harris \& Harris 2000 and paper I).
For consistency with previous work, we use the evolutionary tracks for
0.8$M_{\odot}$ red giants by \citet{van00}, calibrated to $(V-I)$
colors and normalized to standard Milky Way globular cluster fiducial
sequences as discussed in more detail by \citet{hh02}.  To extend the
\citet{van00} grid to the most metal-rich stars ([m/H] $> -0.4$) we
use empirical tracks from two metal-rich star clusters (NGC 6553,
6791) as discussed in \citet{hh02}.

In Figure 6 we have re-plotted the $\mathcal{M}3$ CMD, overlaid with
the model evolutionary tracks.  The model grid was shifted by the M31
distance moduli ($(m-M)_I=24.59$ and $(m-M)_I=24.58$) and reddenings
($E(V-I) = 0.08$ and $0.07$) for the $\mathcal{R}1$ and $\mathcal{M}3$
fields, respectively (see Fig.~8 of Paper I for the CMD of the
background $\mathcal{R}1$ with superimposed tracks.  The excess
population of RGB stars in the $\mathcal{M}3$ field at intermediate
colors and fainter than $I=20.5$ is evident; eg. Figure 2).

For each star in the $\mathcal{R}1$ and $\mathcal{M}3$ CMDs fainter
than the RGB tip ($I\sim 20.5$) we first derive $M_{bol}$ and
$(V-I)_0$, and then interpolate within the model grid in the
($M_{bol}, (V-I)_o$) plane to estimate the heavy-element abundance $Z$
(and [m/H]).  Then, to construct the MDF, we use all stars in the
range $-3.5 < M_{bol} < -1.8$\footnote{Our faint-end cutoff at
$M_{bol} = -1.8$ is slightly brighter than that adopted for the
$\mathcal{M}2$ field in paper I because of the limiting-magnitude
differences discussed above.} to minimize the effects of photometric
spread in the $I$ band and the rather large contribution of blue
objects (likely faint, unresolved galaxies) fainter than $I=22.5$.

Finally, the incompleteness-corrected number of stars $N_c$ (where
$N_c = \sum f(I,V-I)^{-1})$ was derived for 0.1-dex metallicity bins,
and the cleaned MDF of our field was constructed by simply subtracting
the background MDF of $\mathcal{R}1$ from the raw $\mathcal{M}3$ MDF.
This final MDF is plotted in Figure 7 and listed in the first 2
columns of Table 2.  The quoted uncertainties in each metallicity bin
are the Poisson errors in the background-subtracted totals ($\pm
\sqrt{N({\mathcal{M}3 + \mathcal{R}1})}$). In Fig.~7, we also plot
representative errorbars in the derived metallicities that would be
generated solely by measurement uncertainties in $(V-I)$. At the blue
(extreme metal-poor) end of the distribution, the RGB tracks change
very little with metallicity, and even small color errors unavoidably
translate to a large spread in metal abundance.

Nominally, the deeper $V$ limit of the 30 kpc field (compared to that
of the 20 kpc field) should allow us to quantify our MDF to slightly
higher metallicities, since it is the $V$ band that determines the MDF
limit for the extremely red stars with Solar metallicities and above.
However, there appear to be very few such stars in the sample in any
case \citep[see also the HST studies by][]{hol96,rich96a,brown03}, and
their small numbers are additionally uncertain because of the
background contamination.  We thus restrict our MDF to sub-Solar
metallicities.

\subsection{Comparison with 20 kpc field}

Inspection of Figure 7 shows that the M31 outer-halo stars span a wide
range in metal abundance ($-2 < $[m/H]$ < 0$).  The MDF as a whole is
predominantly metal-rich (peaking around [m/H]$\sim -0.5$) but has a
significant metal-poor tail (perhaps with a secondary peak at
[m/H]$\sim -1.5$).  The {\it median} metallicity of the entire MDF is
[m/H]$\sim -0.7$ or [Fe/H]$\sim -1.0$.  All these features are closely
similar to those from the 20 kpc field of Paper I.

Since the publication of paper I, the fiducial sequences for the
metal-rich end of the distribution have been refined somewhat
\citep{hh02}, and we have also made minor changes to our bilinear
interpolation routines.  Accordingly, to make the two MDFs as strictly
comparable as possible, we have re-created the $\mathcal{M}2$ MDF
using the new interpolation routine and exactly the same range in
$M_{bol}$ as used above for the $\mathcal{M}3$ field.  This revised
MDF is listed in Table 2 and plotted in Figure 8, along with the
original MDF from paper I.  As expected, the MDF is virtually
identical (after allowing for a different number of stars as a result
of the more conservative magnitude range adopted here), showing that
the modifications have no significant effect on the discussion from
Paper I.

We show our final direct comparison between the 20 kpc and 30 kpc
fields in Figure 9.  Here, the number of $\mathcal{M}2$ stars has been
multiplied by a factor of 0.23 to normalize both fields to the same
effective population.  It is evident that little difference exists
between the two MDFs.  A formal $\chi^2$ test over the range $-2.25 <$
[m/H] $< -0.15$ indicates that they are similar at the 27\% confidence
level.  In fact, any formal statistical difference is almost entirely
due to the relative excess of stars in only 2 of the 21 bins (each of
which in turn is based on a very small number of stars).  For all the
other bins, the error bars of the two MDFs mutually overlap.

\subsection{Spatial Distribution}

A by-product of our MDF comparison is a limited investigation of the
radial profile of the outer M31 halo.  \citet[][hereafter
PvdB94]{pv94} combined measurements from various sources to construct
the surface brightness profile of the M31 halo, finding that the
profile could be fit by an r$^{1/4}$ law from 0.2 to 20 kpc.  The
outer halo alone could be approximated by a steep power law (stellar
volume density $\rho \propto r^{-4}$ to r$^{-5}$).  As we have not
independently derived the absolute values of $\mu_V$ from our data
(such an analysis is beyond the scope of this paper), we limit
ourselves to a deriving the relative number densities for the
$\mathcal{M}2$ and $\mathcal{M}3$ fields.

The magnitude range for the stars in both MDFs is similar, so the
normalization of the MDFs should directly yield the relative number
density. The total background-corrected number counts in each MDF (up
to [m/H]$=-0.15$) in Table 2 are $N(\mathcal{M}3)=486\pm 45$ and
$N(\mathcal{M}2)=2084\pm 65$, making the number ratio
$N(\mathcal{M}2)/N(\mathcal{M}3) = 4.3\pm 0.6$.  Converted to a
surface brightness ratio, this means the $\mathcal{M}3$ field is
$1.60\pm 0.15$ magnitudes fainter than $\mathcal{M}2$. If we adopt the
(admittedly uncertain) surface brightness of $\mu_V =
29.35^{+1.19}_{-0.55}$ for the 20 kpc field from PvdB94, this suggests
the observed surface brightness of our 30 kpc field is $\mu_V\sim 31
\pm 1$.  This shows the advantage (and in fact, the necessity) of
using a large field size to detect extremely low surface brightness
through direct starcounts; in the small fields used by PvdB94, the M31
halo RGB was barely detected in the $\mathcal{M}2$ field, and only an
upper limit for $\mathcal{M}3$ could be derived ($\mu_V> 30.3$,
consistent with our new result).

Assuming the $r^{1/4}$ law profile from PvdB94 and a minor axis
effective radius $r_e$ of 1.4 kpc (we have scaled the $r_e = 1.3$ kpc
value from PvdB94 for a distance of 780 kpc), we predict
$N(\mathcal{M}2)/N(\mathcal{M}3)$ in the range 4.7 to 5.6 (assuming an
error of $\pm 0.1$ in $r_e$ and an error of $\pm 2$ kpc in the centers
of the $r=20$ kpc and $r=30$ kpc bins).  Thus the stellar density in
the $\mathcal{M}3$ field is {\sl consistent with the M31 spheroid
continuing as an $r^{1/4}$ profile out to at least $r=30$ kpc}.

Similarly, we can derive the stellar surface density decrease from $r=20$ kpc
to $r=30$ kpc as a power law, from which we derive $\sigma \propto r^{-3.6\pm
0.4}$, or a volume stellar density $\rho \propto r^{-4.6\pm 0.4}$; these
numbers are, again, consistent with power law values derived by PvdB94 for $10
\leq r \leq 20$ kpc.  Note, however, that we are unable to discriminate between
the two different profiles with our results.   Studies of the stellar density
of stars in fields at still larger radii will be required to fully address this
issue (e.g. J. Ostheimer et al., in preparation).

\section{Discussion}

The primary results from our analysis are that

\noindent (a) {\it the outer halo of M31 is
predominantly broad and metal-rich to arbitrarily large galactocentric distances}, and

\noindent (b) {\it the outer halo shows little or no radial metallicity gradient}.

In turn, our observed MDFs are (for the most part) very similar to
those found by other researchers in more interior fields
\citep{dur94,hol96,rich96a,reit98,bell03,brown03}.  This statement does not
contradict the recent findings of significant substructure in the M31 halo
\citep{ferg02,mor03,zucker04}, as many of the halo fields are not located near the
stream and warped disk features found in the \citet{ib01} survey.  It
appears therefore that many of the MDF studies to date have been
sampling the `typical halo' of M31.  The contributions of accretion of
satellite galaxies do, however, appear to be an important (although
not necessarily the most {\it dominant}) factor, based on the recent
studies of \citet{ib01} and \citet{ferg02} and the presence of a
significant ``tail'' of metal-poor stars in the MDFs we measure.

The MDF that we and other authors find for the M31 halo, to very large
galactocentric distances, gives no indication that it has a classic
Milky-Way-like metal-poor halo dominating anywhere.
\citet{bell03} find that a more metal-rich {\sl bulge} population begins
clearly within $r < 5$ kpc, but at larger radii they see no
significant increase in the relative size of the metal-poor population
out to $r= 15$ kpc.  Our work now extends this limit to almost 35 kpc.
The combined data make it plausible to suggest that M31 is surrounded
by a predominantly metal-rich, r$^{1/4}$ {\it spheroid}, not unlike
the outer parts of giant elliptical galaxies.

As noted above, we did not assume the age distribution derived for an
inner-halo 11 kpc field by \citet{brown03}.   In their study, they find
that the metal-poor population is classically ``old'' (like Milky Way
globular clusters), but the more metal-rich population is of
intermediate age ($\sim 6-10$ Gyr).  Including some version of this
age-vs.-metallicity gradient in our analysis would clearly yield a
modified MDF which would be still broader and extend to still higher
maximum metallicity, though not in a major way (slightly lower age
would require higher [m/H] to yield the same $(V-I)$ color).

Deep enough photometry to reach the main sequence turnoff is as yet
available for only this one M31 field, and so it is currently unknown
if such a younger stellar population is present at the much larger
halo distances that we have sampled.  Possibilities exist that the
\citet{brown03} sample may include some contamination from a disk
population \citep{rich03}, or from another stellar stream
\citep{gr02}, or something akin to a M31 `thick-disk' population
\citep{wg88,wyse02}.  However, it should be noted that the \citet{brown03}
field is located further from the M31 nucleus than those of some other studies
\citep[eg.][]{dur94,bell03} which found an MDF from the red giants that is {\it
very similar} to that found in our distant fields (see Brown 2003 for a similar
discussion).  Furthermore, none of these MDFs show the additional very red
population observed by \citet{svd01} which clearly could be sampling a thick
disk (see Figure 1).  In addition, \citet{brown03b} has shown that even a
warped M31 disk cannot explain the large fraction of intermediate-age stars in
their study.

The kinematics of these stellar populations in the outer environs of M31 will
be of extreme importance in finally determining their true nature.   If the
flattened M31 `halo' (clearly shown in the maps of \citet{ib01} and
\citet{ferg02}) {\it is} rotation-dominated, it would not be like any thick
disk currently known.  \citet{hk04} have performed a kinematical study
of the planetary nebulae in the outer regions of M31, and clearly show
that the outer parts of the M31 bulge (the same region that we have
called `halo' in this paper) are indeed rotating.  This is suggestive
of either a dominant `bulge' population (if a kinematically distinct
[more metal-poor?] population does exist), or at the very least a
rotating metal-rich `spheroid' (but not a `thick-disk') population.
In addition to this, the outer halos of some giant ellipticals have
significant rotation signatures
\citep{cote01,cote03,peng04a,peng04b}. For all of the differences between the
M31 and Milky Way halos emphasized above, the Milky Way may also have a
somewhat flattened outer stellar halo \citep[eg.][and references
within]{mor00,yanny00,chen01,siegel02}.

We found in Paper I that the observed MDF of our 20-kpc field could be
well matched by a single-zone simple chemical evolution model with a
numerical `yield' $y_{eff} \simeq 0.3 Z_{\odot}$.  Though this
effective nucleosynthetic yield is more than 5 times larger than the
best-fit value for the Milky Way halo, it is still low enough to
suggest that almost half the original gas in the region was ejected or
unused in the rounds of star formation which formed these stars.  By
inference, we can make the same suggestion about the 30 kpc field MDF
since it is virtually identical.  Whether the `proto-M31' region was
created from many dwarf-sized fragments coming together, or the
merging of a few larger, more massive gaseous protogalaxies cannot be
told from these simple chemical models.

Other mechanisms to form metal-rich spheroid populations have also
been explored. In the dissipationless-accretion model of
\citet{cote00}, the MDF of the halo is generated from multiple
accretions of satellites which have already formed most of their
stars, although the mass spectrum of such satellite dwarf galaxies
would be required to be strongly weighted towards the high-mass end if
the MDF is as metal-rich as we see in M31 and in large ellipticals
\citep{hh01}. By contrast, \citet{beas03} use CDM-based hierarchical 
merging models of gaseous fragments to construct MDFs that are
typically broad and metal-rich, closely resembling what we see in M31
or (e.g.) NGC 5128.  In this model, the sequence of mergers typically
continues for a few Gyr and so the metal-richer population can easily
be younger by a few Gyr than the first, metal-poor population. Lastly,
\citet{bek03} explored the possibility that the merger of two disk
galaxies could explain the NGC 5128 halo MDF, and further suggested
that the M31 halo could have formed via a similar route.  In this
latter model, the stars now in the halo originated in the moderately
metal-rich progenitor disks before the merger.  An advantage of this
disk-merger model is that is can easily explain the fact that the MDF
of the {\sl globular clusters} in the merged galaxy is different from
the halo-star MDF (the metal-poor clusters were the ones already
present in the progenitor galaxies, while most of the metal-richer
ones formed in the merger).  On the other hand, its disadvantages are
that it predicts a modest halo metallicity gradient (which we do not
see in M31), and it does not yet address the necessary later
``reconstruction'' of a major disk in M31.

In summary, at present there are a variety of possible modelling
approaches capable of generating the rather metal-rich halo MDFs that
are now realized to be common in large galaxies.  However, the strong
similarities of the halo MDFs to date, the presence of metallicity
gradients in the inner spheroids, the small metal-poor population at
any radius, and the spheroid luminosity-metallicity correlation
\citep[][]{rich04} all suggest that stellar satellite accretion is not
the {\it dominant} factor in spheroid formation for many galaxies.

\section{Summary}

We have presented $(I,V-I)$ photometry of the M31 halo within a field
located 30 kpc from the galaxy center along the SE minor axis. Using
careful object classification and statistical removal of background
contamination, we have been able to isolate a sample of a few hundred
M31 halo red giants. We derive the photometric metallicity
distribution function for this outer M31 halo, and find that it is
virtually identical with the MDF from our previously studied 20-kpc
field: the implication is that the MDF throughout the halo is broad,
moderately metal-rich (peaking at [m/H] $\simeq -0.5$), and has no
systematic radial gradient.  We also find that the M31 halo has a
steep radial density profile: the elliptical-like $r^{1/4}$ profile
determined previously from data at at smaller radii continues on
outward through our 30-kpc field.

\acknowledgments

The authors would like to thank Denise Hurley-Keller, Pat
C\^{o}t\'{e}, Steve Majewski, Mike Rich, Jamie Ostheimer and Raja
Guhathakurta for numerous related discussions, and David Bohlender for
assistance with the UH8K archival data.  PRD gratefully acknowledges financial
support from Robin Ciardullo.  We acknowledge use of facilities made
available by the Canadian Astronomy Data Centre, which is operated by
the Herzberg Institute of Astrophysics, National Research Council of
Canada.  WEH and CJP acknowledge support from the Natural Sciences and
Engineering Research Council of Canada.


\clearpage

\begin{deluxetable}{ccc}
\tablenum{1} \tablewidth{0pt} \tablecaption{Limiting Magnitudes} \tablehead{
\colhead{CCD} & \colhead{$V_{lim}$} & \colhead{$I_{lim}$} } \startdata
0 &  24.65 & 23.51 \\
1 &  24.78 & 23.61 \\
2 &  24.61 & 23.38 \\
3 &  24.76 & 23.62 \\
5 &  24.71 & 23.65 \\
7 &  24.56 & 23.57 \\
\enddata
\end{deluxetable}

\clearpage

\begin{deluxetable}{rcrrcrr}
\tablenum{2} \tablewidth{0pt} \tablecaption{Metallicity Distribution
Functions\tablenotemark{a}} \tablehead{ \colhead{} & \colhead{}&
\multicolumn{2}{c}{$\mathcal{M}2$\tablenotemark{b}} & \colhead{} &
\multicolumn{2}{c}{$\mathcal{M}3$\tablenotemark{b}}\\
\cline{3-4} \cline{6-7}\\
\colhead{[m/H]} & \colhead{} &\colhead{$N_c$} & \colhead{$\sigma$}& \colhead{}
&\colhead{$N_c$} & \colhead{$\sigma$} \\  } \startdata
$-2.2$ &&     6.4  &   2.6 &&     1.1  &   1.1   \\
$-2.1$&&     12.9  &   5.0 &&     2.5 &    3.7  \\
$-2.0$ &&    12.8  &   5.2  && $-$1.9 &    3.3   \\
$-1.9$ &&    41.8  &   7.4  &&    3.6 &    3.9   \\
$-1.8$&&     24.1  &   5.8 &&     1.4 &    3.2   \\
$-1.7$&&     38.8  &   6.7  &&   14.2 &    4.5   \\
$-1.6$&&     28.4  &   6.2  &&   17.5 &    5.3  \\
$-1.5$ &&    48.3  &   8.7  &&   14.9 &    6.5 \\
$-1.4$ &&    50.8  &   8.4   &&  12.7 &    5.6  \\
$-1.3$ &&    60.3  &   9.1  &&   10.5 &    5.6 \\
$-1.2$ &&    58.0  &   9.5  &&   33.4 &    8.1  \\
$-1.1$ &&   108.5  &  11.8 &&    22.7 &    7.0 \\
$-1.0$ &&    96.7  &  11.8  &&   21.9 &    7.9  \\
$-0.9$ &&   164.3  &  14.8  &&   33.7 &    8.6  \\
$-0.8$ &&   164.5  &  14.5 &&    47.2 &    9.2  \\
$-0.7$  &&  196.8  &  17.3  &&   30.1 &   10.5   \\
$-0.6$  &&  262.2  &  21.1 &&    37.3 &   12.6  \\
$-0.5$ &&   257.1  &  21.8 &&    52.5 &   14.2  \\
$-0.4$   && 216.3  &  23.8 &&    38.2 &   17.8  \\
$-0.3$ &&   137.9  &  22.4  &&   44.2 &   18.6  \\
$-0.2$ &&   142.0  &  25.9  &&   49.1 &   19.5   \\
$-0.1$  &&   \nodata  &  \nodata  &&      9.7 &   20.1  \\
\enddata
\tablenotetext{a}{for M31 stars in the range $-3.5 < M_{bol} < -1.8$}
\tablenotetext{b}{counts after background subtraction}
\end{deluxetable}

\clearpage

\begin{figure}
\epsscale{1.00}
\plotone{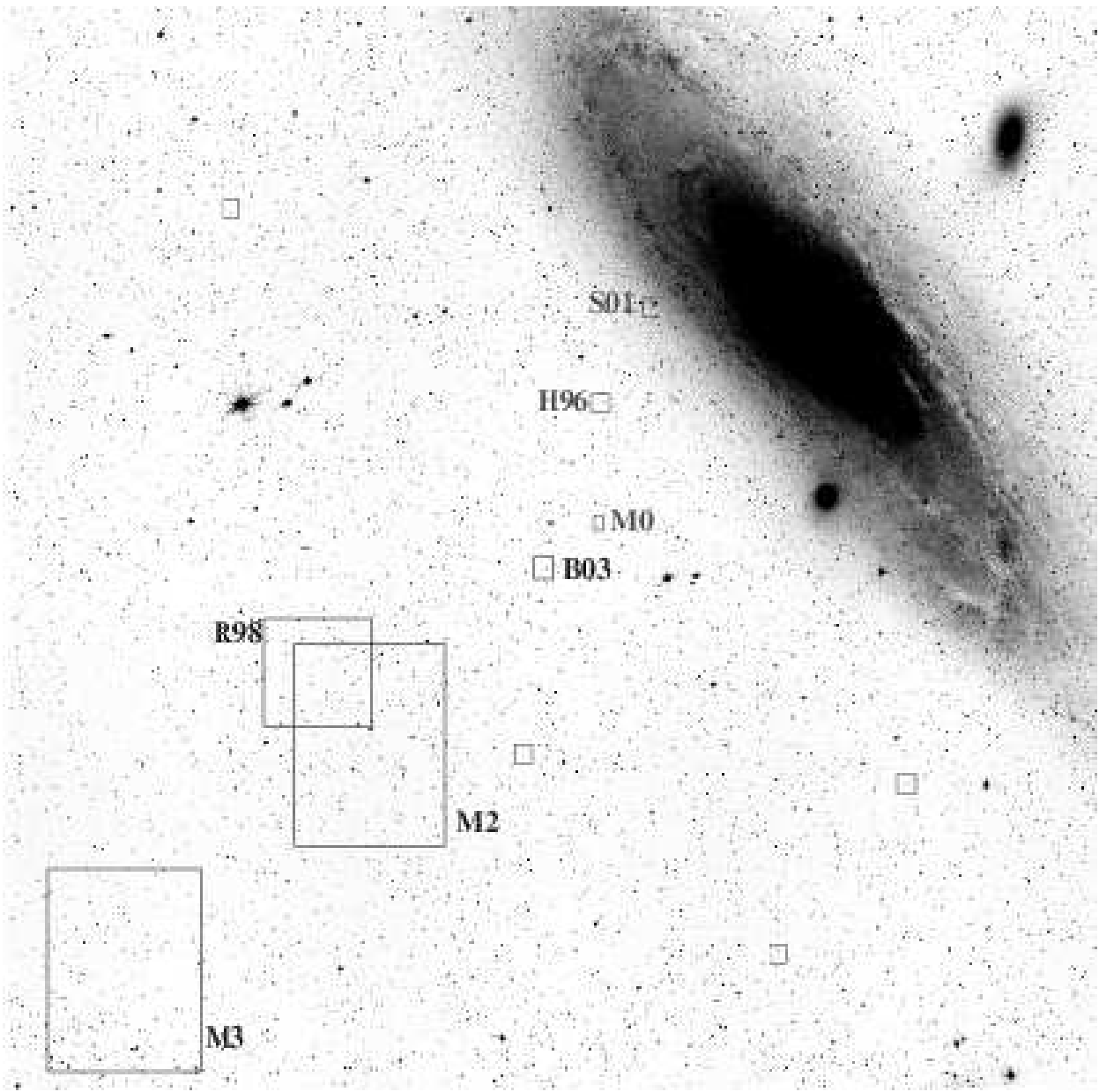}
\caption{Location of the M31 halo field ($\mathcal{M}$3)
under study in this paper -- the field size is that of the usable area of the
UH8K mosaic (21$^\prime$ x 28$^\prime$).  The locations of fields from other
studies are also plotted (with rough field sizes indicated), and labelled as
follows : $\mathcal{M}$2 (20 kpc field; Durrell et al. 2001), R98 (Reitzel et al. 1998),
B03 (Brown et al. 2003, and Holland et al. 1996), H96 (Holland et al. 1996),
S01 (Sarajedini \& van Duyne 2001) and M0 (Durrell et al. 1994).  The 4
unlabelled boxes are some of the halo fields from Belazzini et al. (2003).   The
image is from the Digitized Sky Survey, and is 2\degpoint 5 on a side.}
\label{fig1}
\end{figure}

\begin{figure}
\epsscale{1.00}
\plotone{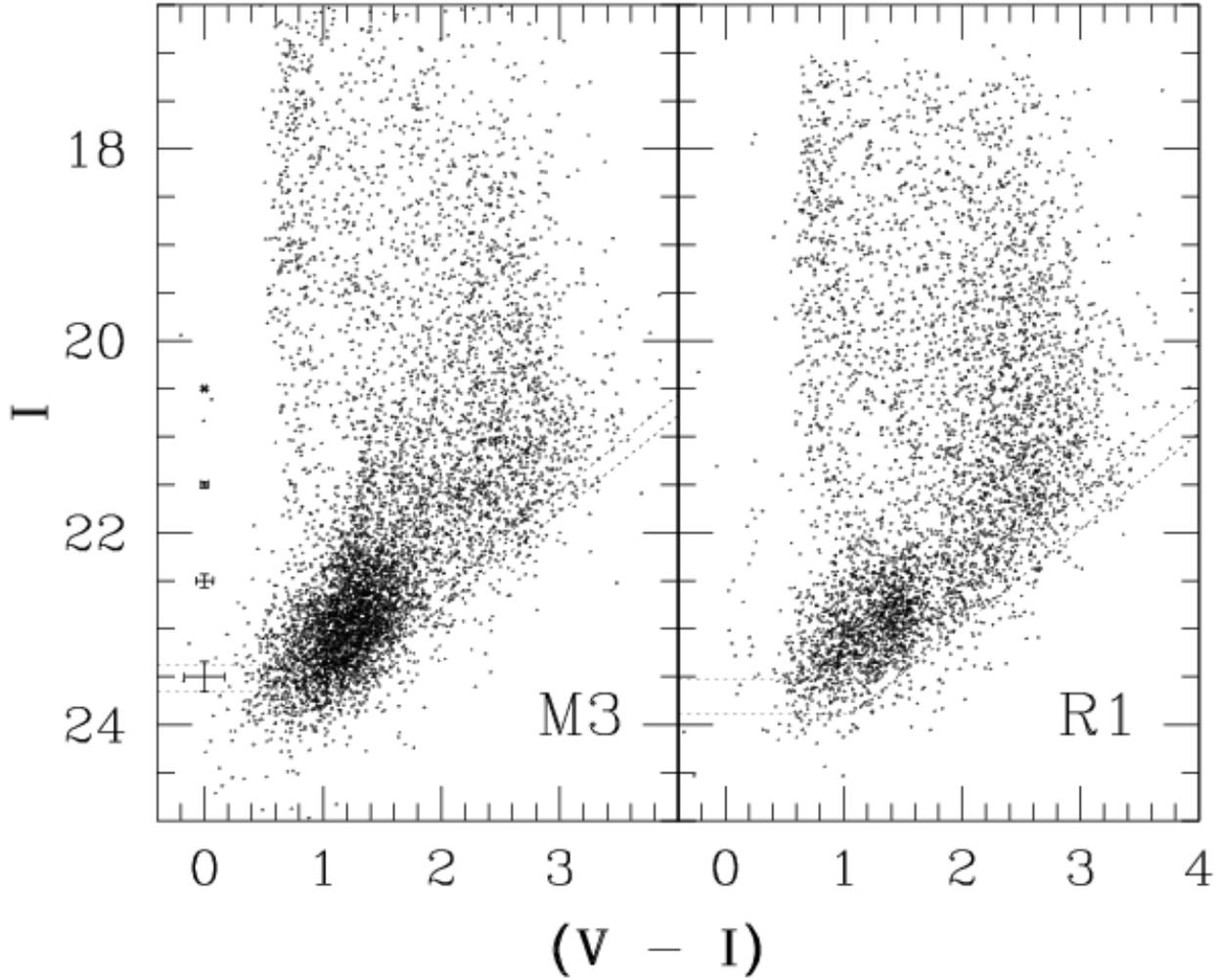}
\caption{$VI$ Color magnitude diagrams of the (left) $\mathcal{M}$3 halo
field and the $\mathcal{R}1$ background field (right), 
based on data from all 6 usable chips of the CCD array.   The dashed
lines denote the full range of the 50$\%$ completeness levels for the CCDs
used.    All non-stellar objects have been rejected via image classification.
The error bars plotted in the $\mathcal{M}3$ CMD denote the {\it representative} 
uncertainties for objects with $(V-I)= 1$.}  
\label{fig2}
\end{figure}

\begin{figure}
\epsscale{1.00}
\plotone{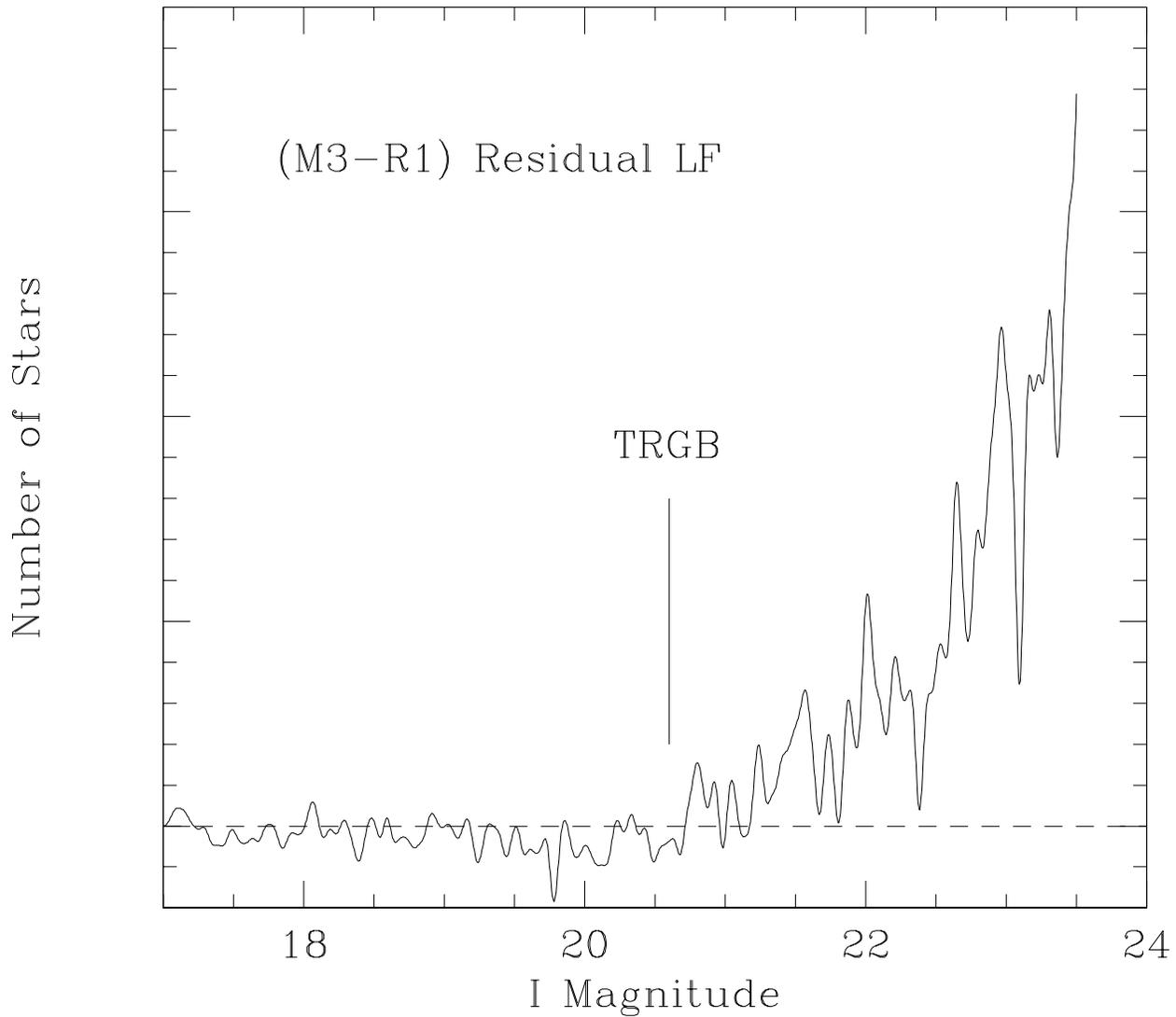}
\caption{Residual $I-$band luminosity function  
of all stellar objects, after subtraction of the
background $\mathcal{R}1$ field from the $\mathcal{M}3$ field.  The
LF has been smoothed with a 0.03 mag kernel.  The
expected location of the RGB tip (from our results in paper I) is
indicated.}
\label{fig3}
\end{figure}

\begin{figure}
\epsscale{1.00}
\plotone{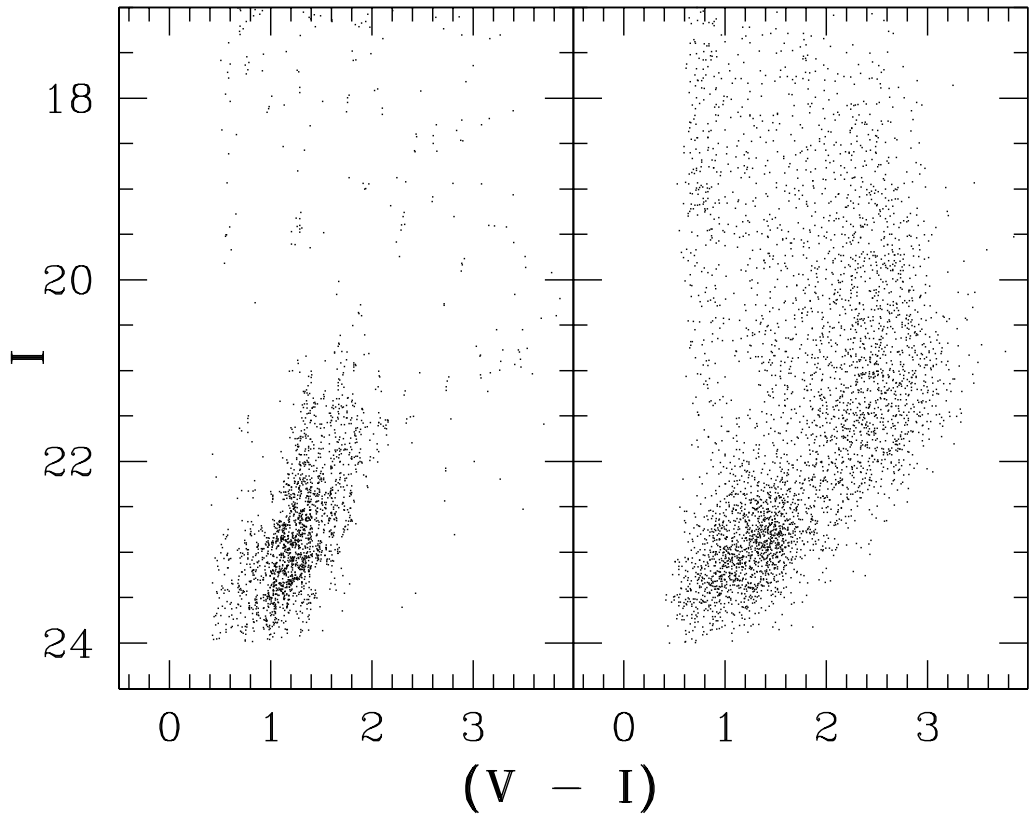}
\caption{Residual color-magnitude diagrams.  (left) `Cleaned' 
color-magnitude diagram for the $\mathcal{M}3$ field, after removal of
objects that matched objects in the background $\mathcal{R}1$ field.
(right) The CMD of all objects that were removed.}
\label{fig4}
\end{figure}

\begin{figure}
\epsscale{1.00}
\plotone{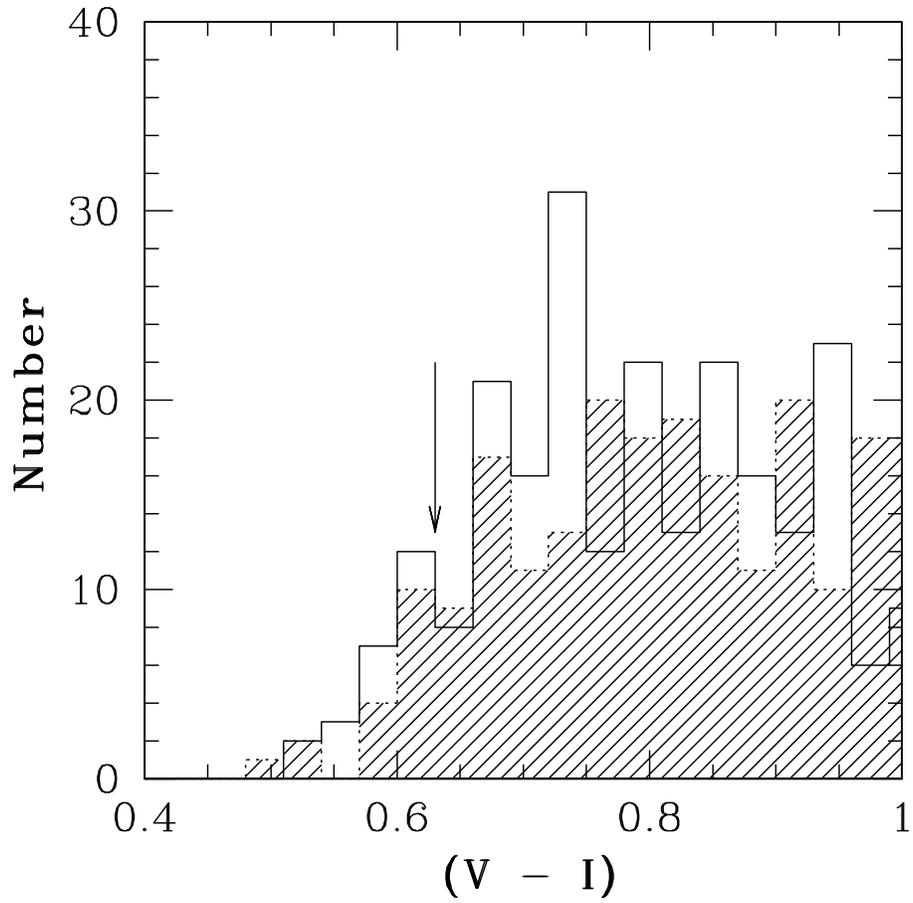}
\caption{(V-I) color distribution of foreground stars
(all objects with $17 < I < 20$) in the $\mathcal{M}2$ (filled histogram) and
$\mathcal{M}3$ (solid line) CMDs.  The $\mathcal{M}3$ data has been shifted
redder by 0.03 mag to match the $\mathcal{M}2$ data, with the location of the
blue edge derived for that dataset (Paper I) labelled as an arrow.}
\label{fig5}
\end{figure}

\begin{figure}
\epsscale{1.00}
\plotone{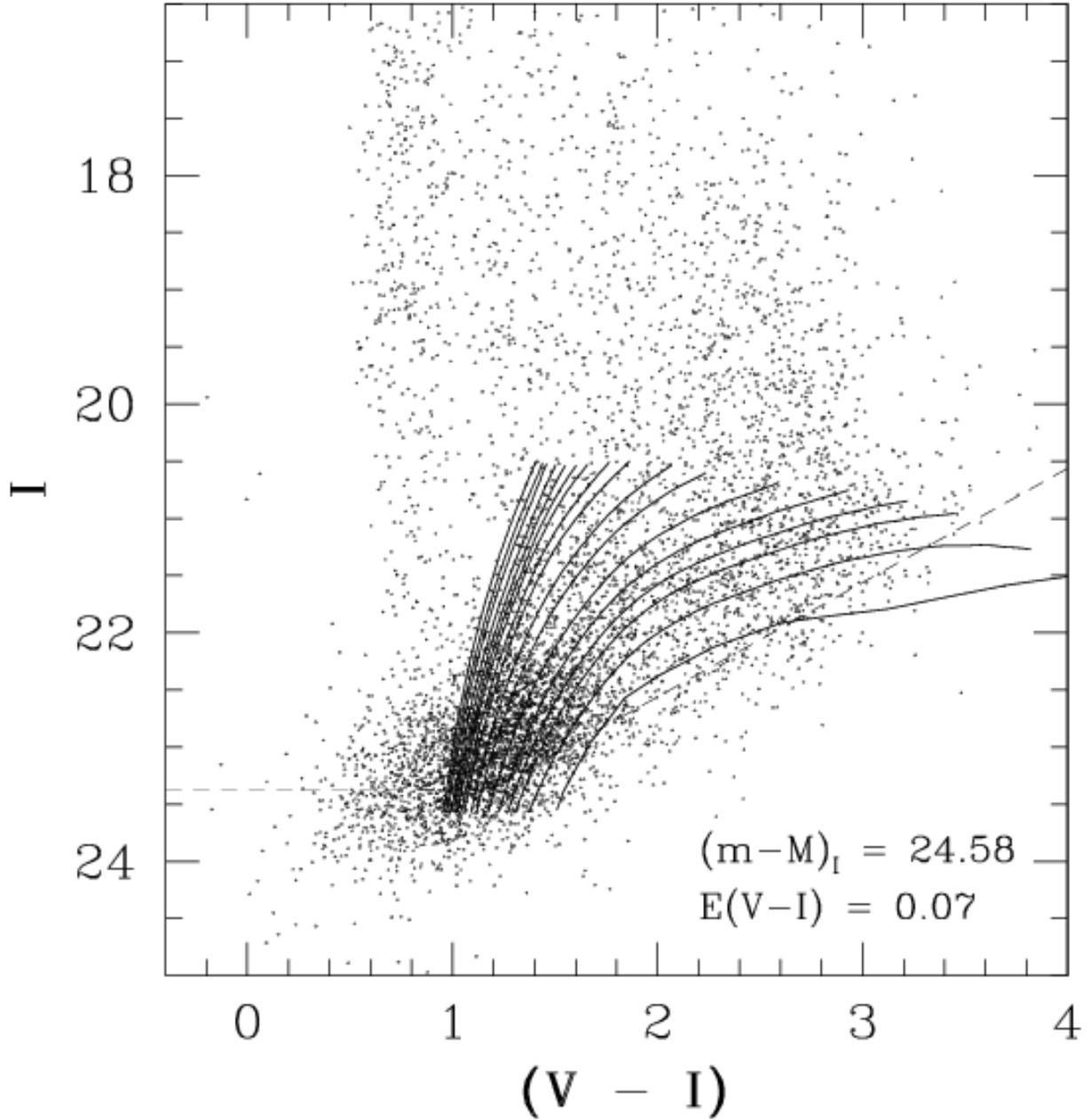}
\caption{CMD for stars in the $\mathcal{M}3$ field.  The
solid lines are evolutionary tracks for 0.8 $M_{\odot}$ stars from VandenBerg
et al. (2000), shifted to the distance modulus and reddening shown.  The models
have been further shifted 0.03 mag to the blue (empirical correction - see paper I
for more details).   From left to right : [Fe/H] = $-2.31$, $-2.14$, $-2.01$,
$-1.84$, $-1.71$, $-1.61$, $-1.54$, $-1.41$, $-1.31$, $-1.14$, $-1.01$,
$-0.83$, $-0.71$, $-0.61$, $-0.53$ and $-0.40$.   The rightmost model is the
[Fe/H]$=+0.07$ 12 Gyr isochrone from Bertelli et al. (1994).  The dashed line
represents the 50$\%$ completeness level for the least-sensitive chip in the
mosaic.}
\label{fig6}
\end{figure}

\begin{figure}
\epsscale{1.00}
\plotone{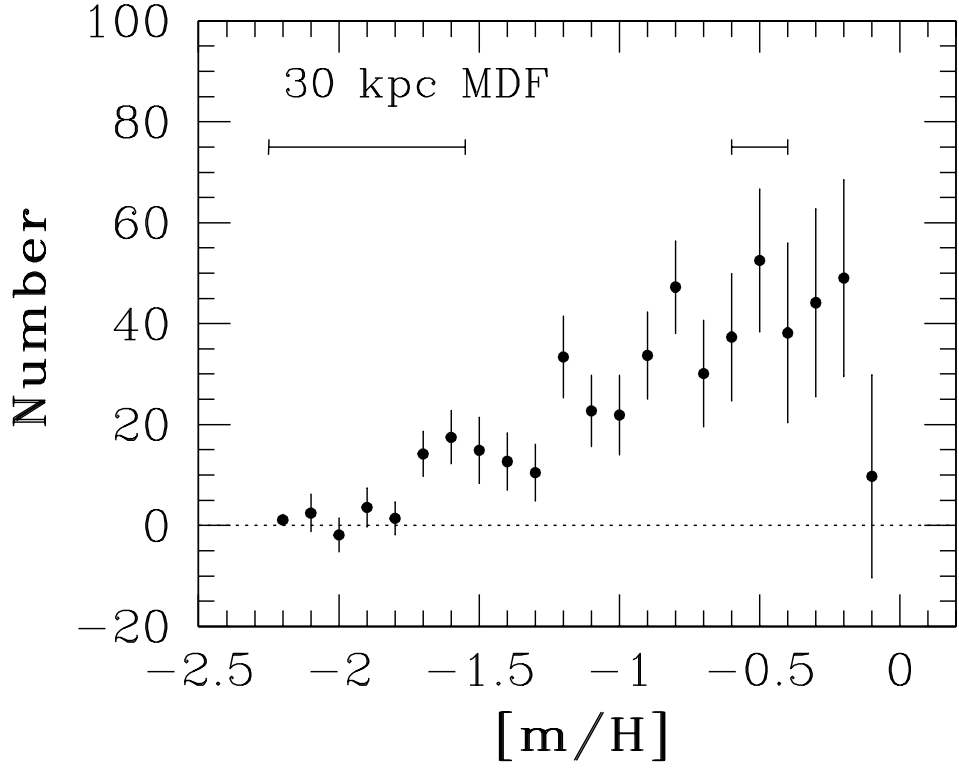}
\caption{Metallicity Distribution Function (MDF) for all stars
with $-3.5 < M_{bol} < -1.8$ in the $\mathcal{M}3$ (30 kpc) field.  The
horizontal error bars represent typical metallicity errors based solely on the
$(V-I)$ color uncertainties.   The data has been corrected for both photometric
completeness and for background contamination; see text for details.}
\label{fig7}
\end{figure}

\begin{figure}
\epsscale{1.00}
\plotone{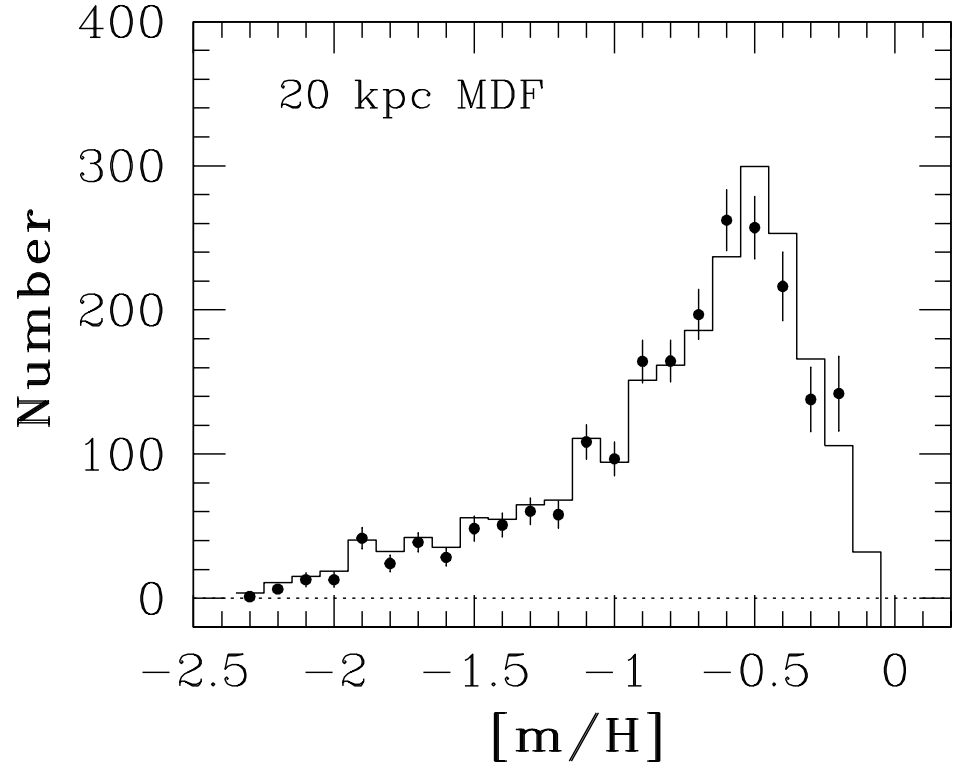}
\caption{Comparison between the MDF of the $\mathcal{M}2$ (20
kpc) field from paper I (solid line) and the revised MDF for the same field
based on the new analysis (filled circles); see text for details.  The paper I
MDF has been arbitrarily scaled by 0.8 to match the new MDF.} 
\label{fig8}
\end{figure}

\begin{figure}
\epsscale{1.00}
\plotone{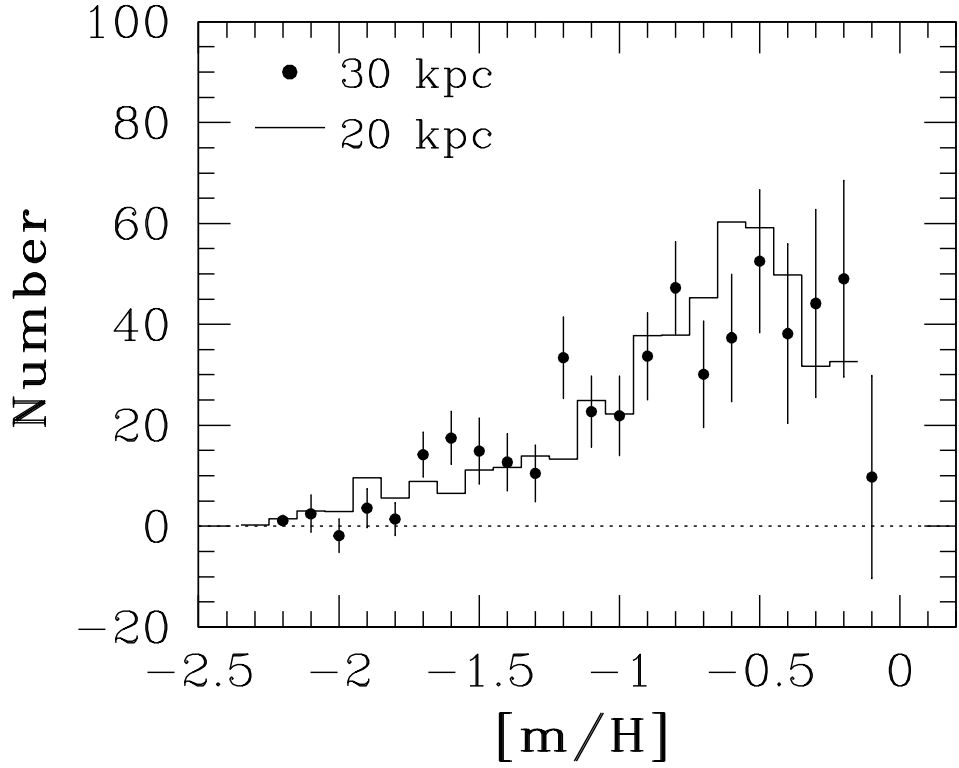}
\caption{Comparison between the MDFs of the 20 kpc field
(solid line) and the 30 kpc field (filled circles).  The 20 kpc MDF has been
scaled by 0.23 to match the total number of objects in the 30 kpc
MDF.}
\label{fig9}
\end{figure}

\end{document}